\documentclass{ws-p8-50x6-00}

\def\GeV{\;\mbox{GeV}}

\usepackage{placeins}

\begin{document}

\title{Inclusive Hadron Production and Dijets at HERA}
\author{Thomas Hadig\\ on behalf of the H1 and ZEUS collaborations}

\address{DESY, Notkestr. 85, 22607 Hamburg, Germany\\E-mail: hadig@rwth-aachen.de}


\maketitle

%{{{}}}
%{{{  Abstract
\abstracts{
This article summarizes a talk given at the International Symposium on
Multiparticle Dynamics 1999 in Providence/USA. It provides an
overview on the variety of measurements of the hadronic final state in
jet production for deep-inelastic scattering and photoproduction at
HERA.}
%}}}
%{{{  Introduction
\section{Introduction}

The HERA storage ring provides beams of protons at $820\GeV$
and positrons at $27.5\GeV$ energy. Collisions of these are recorded
at two general purpose detectors, H1\cite{h1det} and 
ZEUS\cite{zeusdet}. Measurements of the hadronic final state
in deep-inelastic scattering and in photoproduction processes
allow precise tests of the validity of perturbative Quantum-Chromo
Dynamics (pQCD). In addition the extraction of the strong coupling
strength $\alpha_s$ or the parton density functions with high
precision becomes possible.

%}}}
%{{{  Deep-Inelastic Scattering
\section{Deep-Inelastic Scattering}

In deep-inelastic scattering the squared momentum transfer $Q^2$ from
the beam electron to the proton is large. The scattered electron found
in the detector at large scattering angles allows a precise determination of the event kinematics.
The hadronic final state is resolved into jets e.g.\ by means of the
longitudinal-boost invariant $k_t$ algorithm\cite{ktincl}.

\begin{figure}[tb]
\includegraphics[width=.49\hsize,clip]{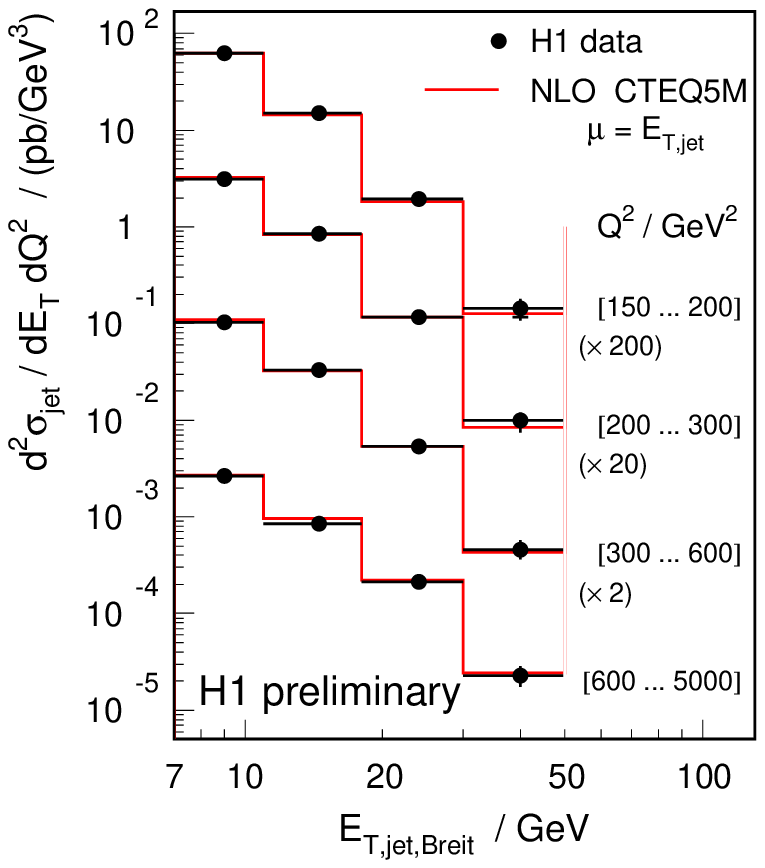}\hspace{.01\hsize}%
\includegraphics[width=.49\hsize,clip]{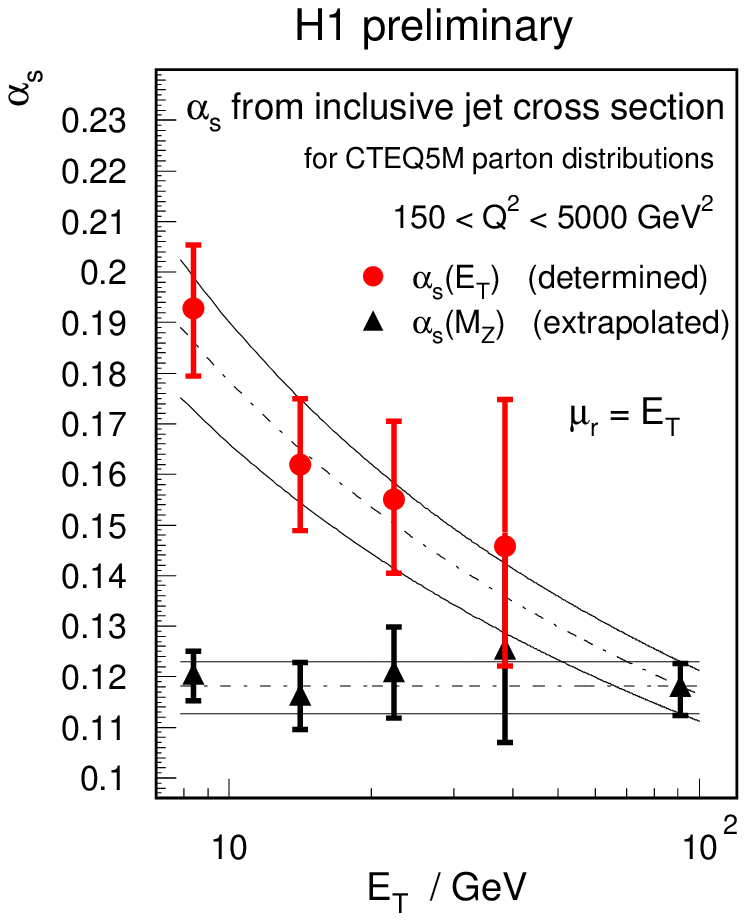}\\
\caption{H1 double differential cross sections for inclusive 
jet production in DIS. The data are compared to NLO
predictions. The strong coupling strength $\alpha_s,$ extracted from the 
inclusive jet distributions, is also shown.}
\label{fig:dis1}
\end{figure}

\begin{figure}[tb]
\includegraphics[width=.45\hsize,clip]{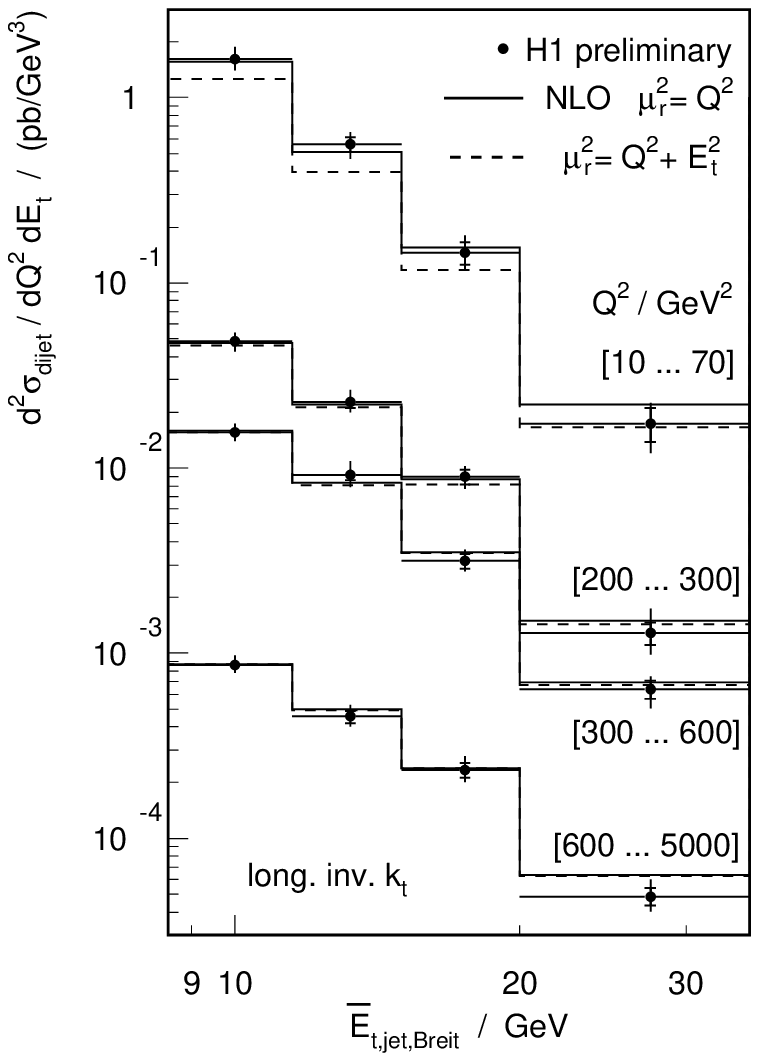}%\hspace{.01\hsize}%
\includegraphics[width=.54\hsize,clip]{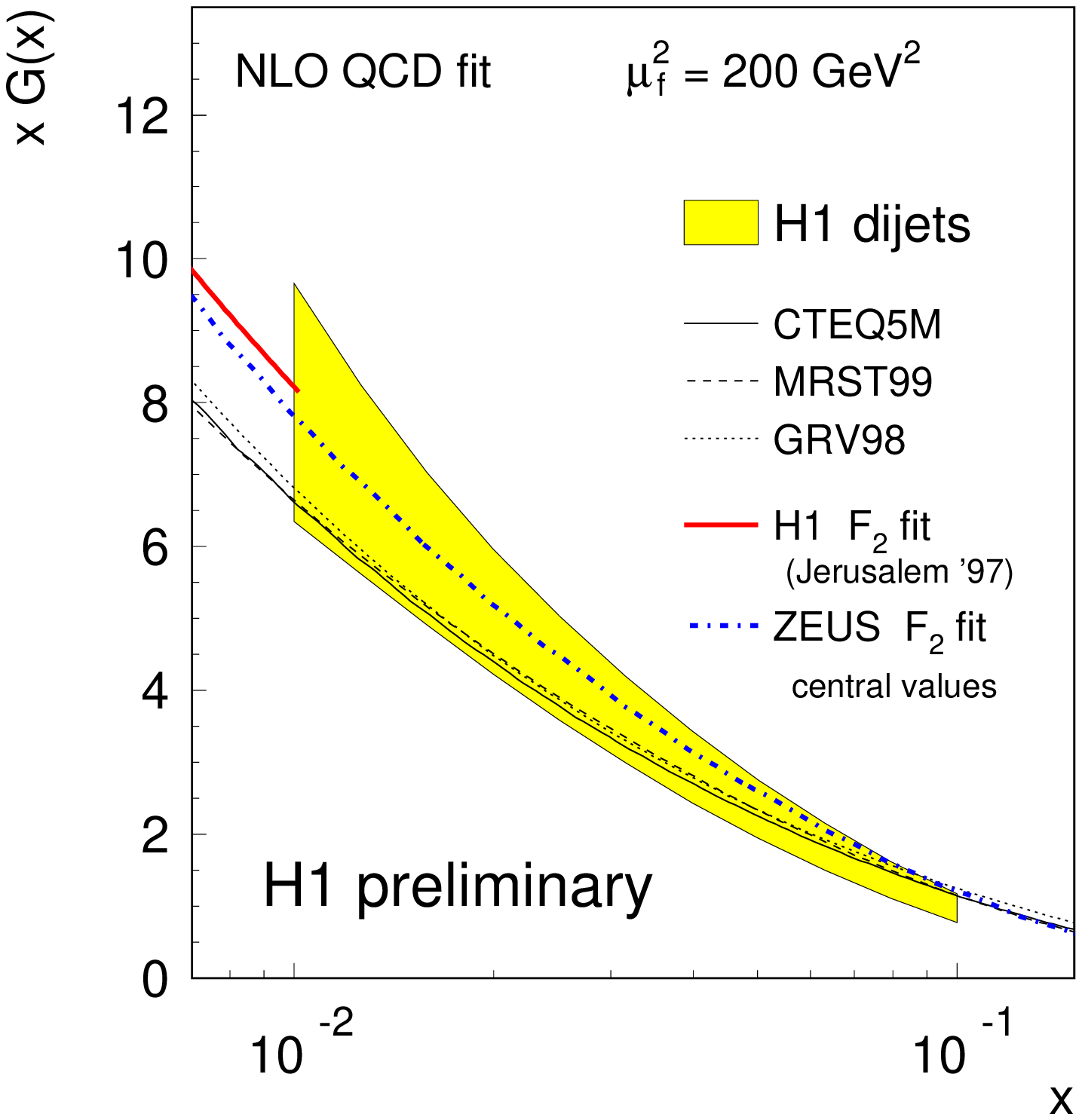}
\caption{H1 double differential cross sections for 
dijet production in DIS. The data are compared to NLO
predictions. The gluon
density, extracted from the dijet
distributions, is also shown.}
\label{fig:dis2}
\end{figure}

Inclusive jet\cite{h1incl}, dijet\cite{h1disdijet,zeusdisdijet}, and
three jet\cite{disthree} cross sections are measured. Inclusive and
dijet distributions are shown in figures~\ref{fig:dis1}
and~\ref{fig:dis2}. NLO pQCD predictions\cite{nloprog} describe the data
well and allow the strong coupling strength $\alpha_s$ to be
extracted. H1 obtained from inclusive jet cross sections
\begin{equation}
\alpha_s(M_Z^2) = 0.1181 \pm 0.0030\mbox{(exp)} ^{+0.0039}_{-0.0046}\mbox{
(theo)} ^{+0.0036}_{-0.0015}\mbox{(pdf)} 
\end{equation}
and ZEUS determined
\begin{equation}
\alpha_s(M_Z^2) = 0.120  \pm 0.003 \mbox{(stat)} ^{+0.005}_{-0.006}\mbox{
(exp)} ^{+0.003}_{-0.002}\mbox{(theo)} 
\end{equation}
from dijet cross sections. Both values are in agreement with the current
world average. The uncertainty on $\alpha_s$ is dominated by the variation of the
hadronic energy scale, the renormalization scale, and the variation of
the parton density functions. The gluon density is determined from dijet
events by H1 as shown in figure~\ref{fig:dis2}. The distribution is in
agreement with the extraction from scaling violations and with global
fits. Three jet cross sections\cite{disthree} are as well measured, but
for the interpretation only LO pQCD calculations are available.

%{{{  figure DIS1
%}}}

%{{{  figure DIS2
%}}}

%}}}
%{{{  Photoproduction 
\section{Photoproduction}

In photoproduction the squared momentum transfer is small $(Q^2\approx 0)$
and the scattered electron most often leaves the experiment undetected.
For jet production the transverse jet energy provides the reference
scale. As for DIS, dijet\cite{photodijet:eps,photodijet:paper} and three
jet\cite{photothree} processes are measured.

\begin{figure}[tbp]
\includegraphics[width=.49\hsize,clip]{zeus_threejet_3body}\hspace{.01\hsize}%
\includegraphics[width=.49\hsize,clip]{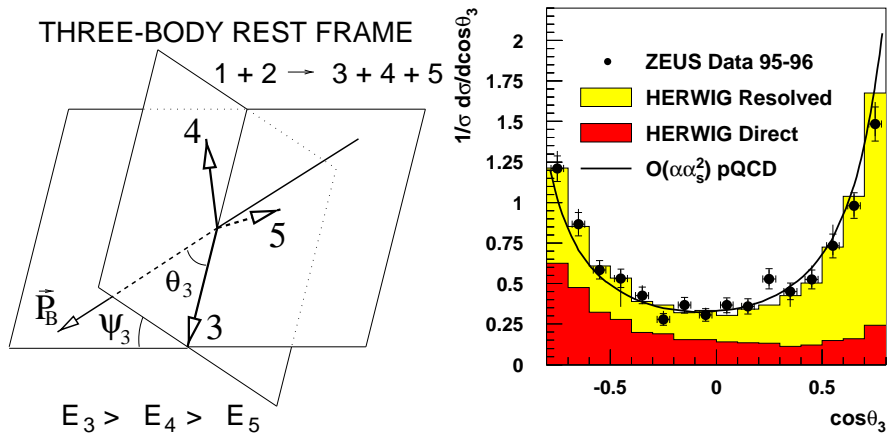}
\caption{The definition and distribution of $\cos(\theta_3)$ for three 
jet events in photoproduction measured by ZEUS. In addition to the data,
the breakdown of the HERWIG Monte Carlo into direct and resolved
component and the theory prediction is shown.}
\label{fig:photo:three}
\end{figure}

In the three jet case the event structure can be described by angular
distributions. Figure~\ref{fig:photo:three} shows the definition and the
distribution of $\cos(\theta_3)$ where $\theta_3$ is the angle of the
jet with the highest energy with respect to the proton beam direction.
The Monte Carlo comparison demonstrates that in addition to the direct
process where the photon directly interacts with the parton, the
resolved process is needed for describing the data. In the resolved
case, the photon fluctuates into a partonic state and one of these
partons interacts with the parton coming from the proton. pQCD
calculations including this process are able to describe the data.

%\FloatBarrier
%{{{  figure
%}}}

\begin{figure}[tbp]
\includegraphics[width=.49\hsize,clip]{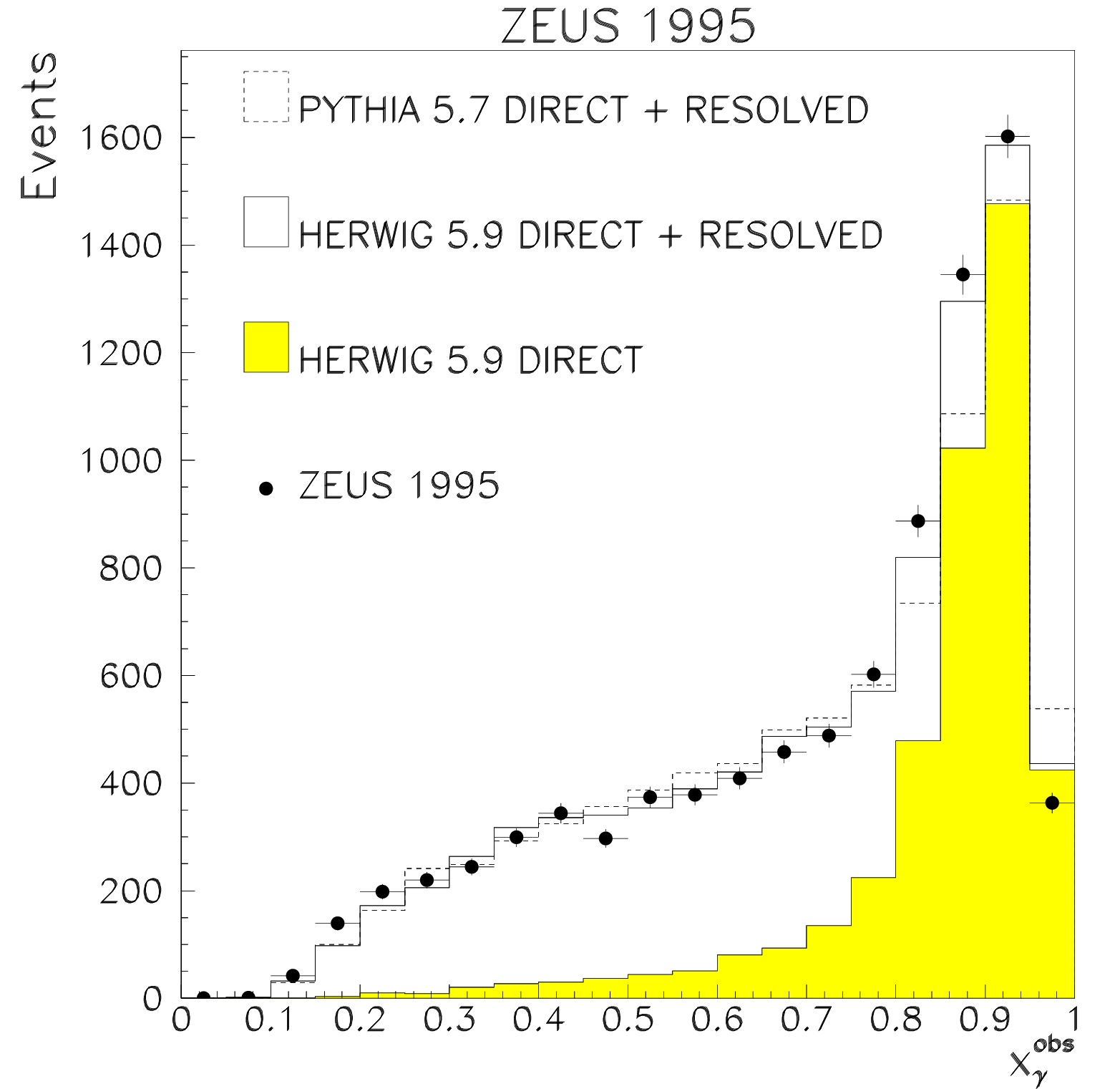}\hspace{.01\hsize}%
\includegraphics[width=.49\hsize,clip]{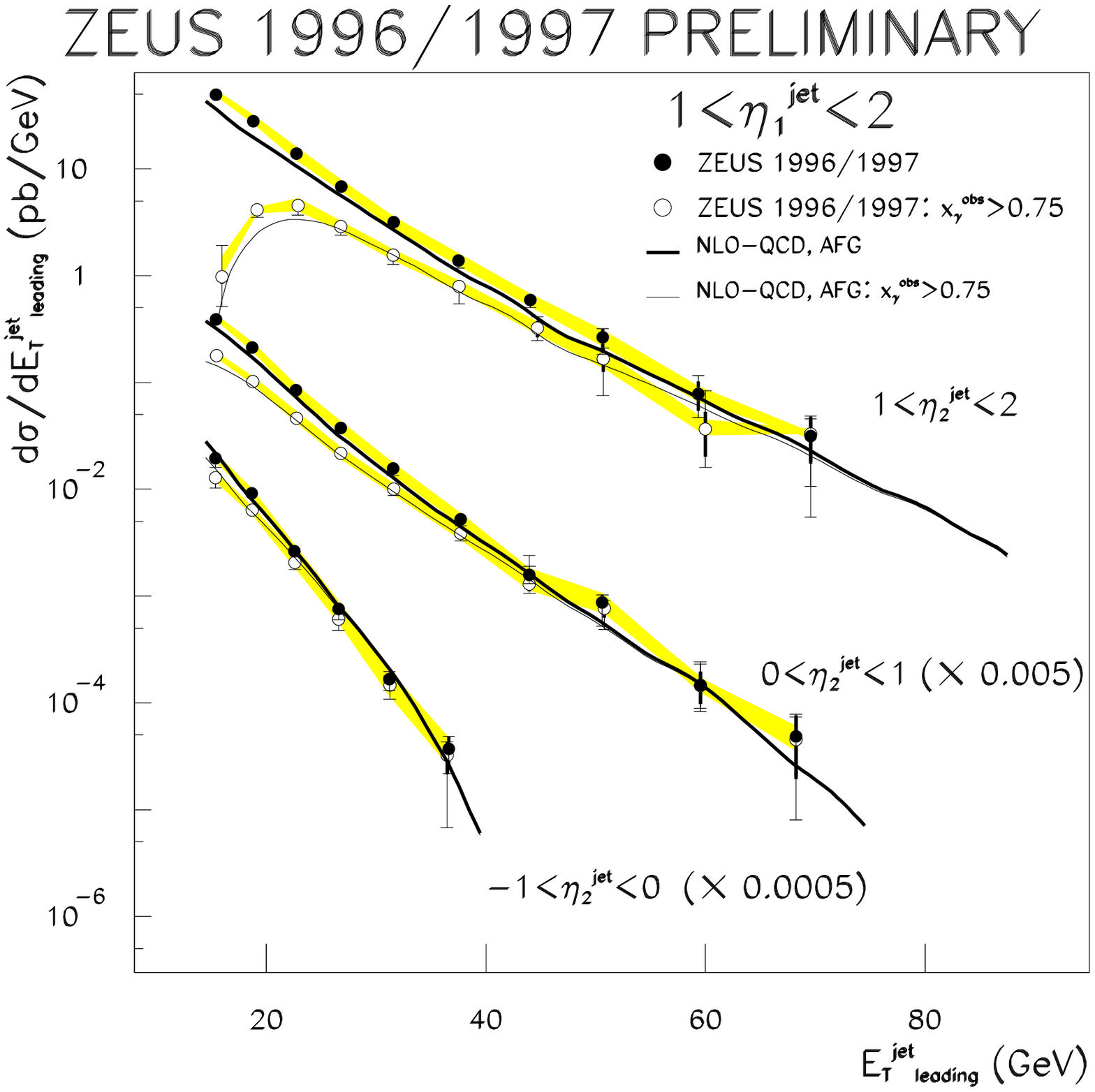}
\caption{$x_{\gamma}^{\mbox{\scriptsize obs}}$ 
spectrum and single differential cross section for dijet events in
photoproduction measured at ZEUS compared to Monte Carlo and NLO pQCD
predictions. The leading (second) jet is required to have a transverse
energy greater than $14 (11)\GeV.$}
\label{fig:photo:dijet}
\end{figure}

\begin{figure}[tb]
\includegraphics[width=.99\hsize,clip]{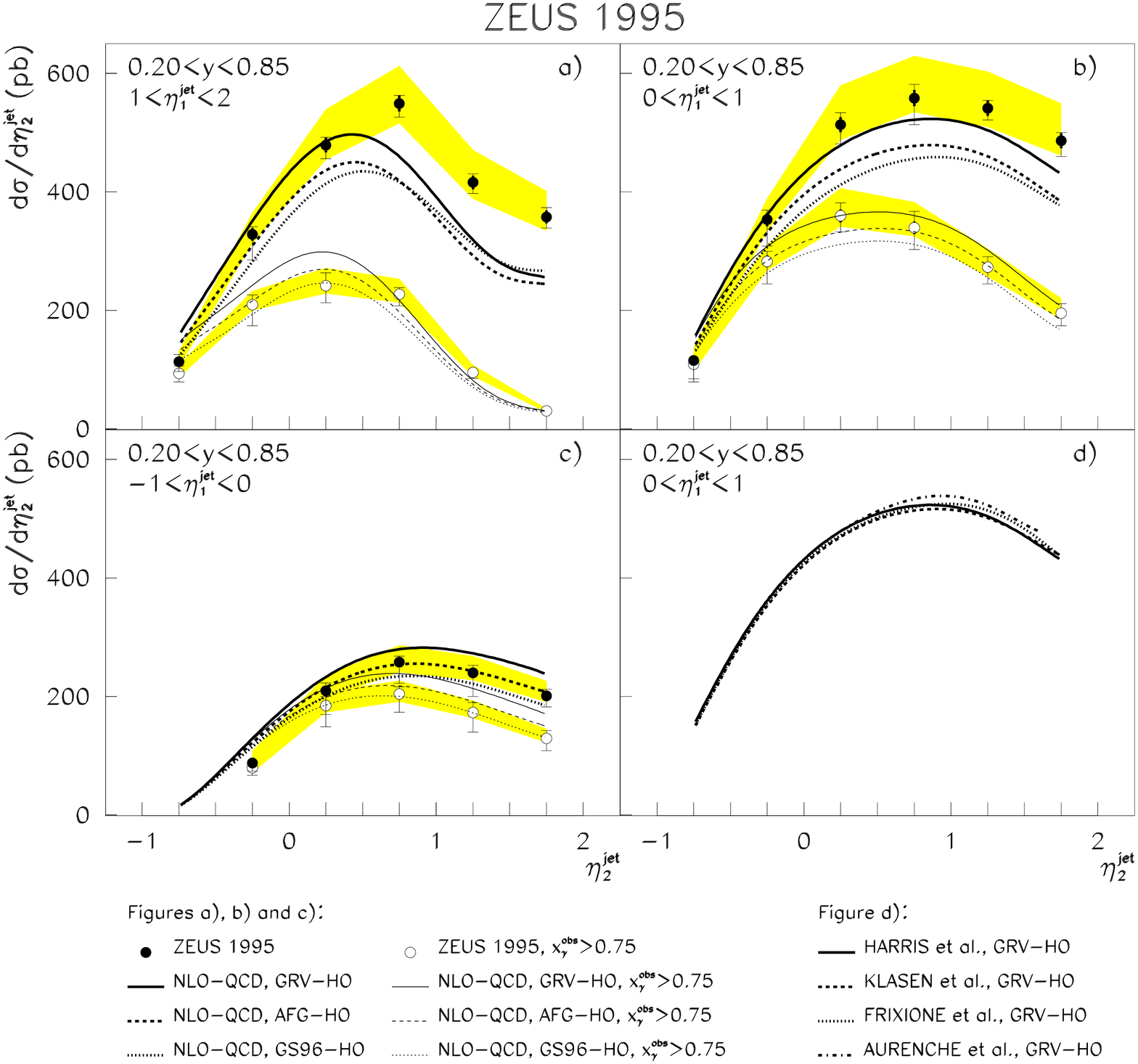}
\caption{
a)-c) Dijet cross section as function of jet rapidity for different
$x_{\gamma}^{\mbox{\scriptsize obs}}$ ranges compared to NLO pQCD
predictions using various photon structure functions. The thick band
shows the systematic uncertainty; d) Comparison of different NLO pQCD
calculations for one choice of photon structure function.}
\label{fig:photo:pdf}
\end{figure}

Figure~\ref{fig:photo:dijet} shows the fraction
$$x_{\gamma}^{\mbox{\scriptsize obs}} = {
E_{T,1}^{\mbox{\scriptsize jet}} e^{-\eta_{1}^{\mbox{\scriptsize jet}}} +
E_{T,1}^{\mbox{\scriptsize jet}} e^{-\eta_{1}^{\mbox{\scriptsize jet}}} \over
2yE_e}$$ 
of the photon momentum observed in dijet processes in the ZEUS data. 
The strong peak close to $x_{\gamma}^{\mbox{\scriptsize obs}}=1$ is 
attributed to direct processes, but
again the resolved component is needed for a description of the data by Monte
Carlo models. NLO pQCD calculations in general describe the cross section
including the dependence on $x_{\gamma}^{\mbox{\scriptsize obs}}$ and the
rapidity of the jets. However, at large $E_{T,\mbox{\scriptsize jet}}$ 
a clear disagreement between data and NLO
prediction for forward jets is found. Figure~\ref{fig:photo:pdf} shows this
more clearly.

%{{{  figure
%}}}

\begin{figure}[tb]
\includegraphics[width=.49\hsize,clip]{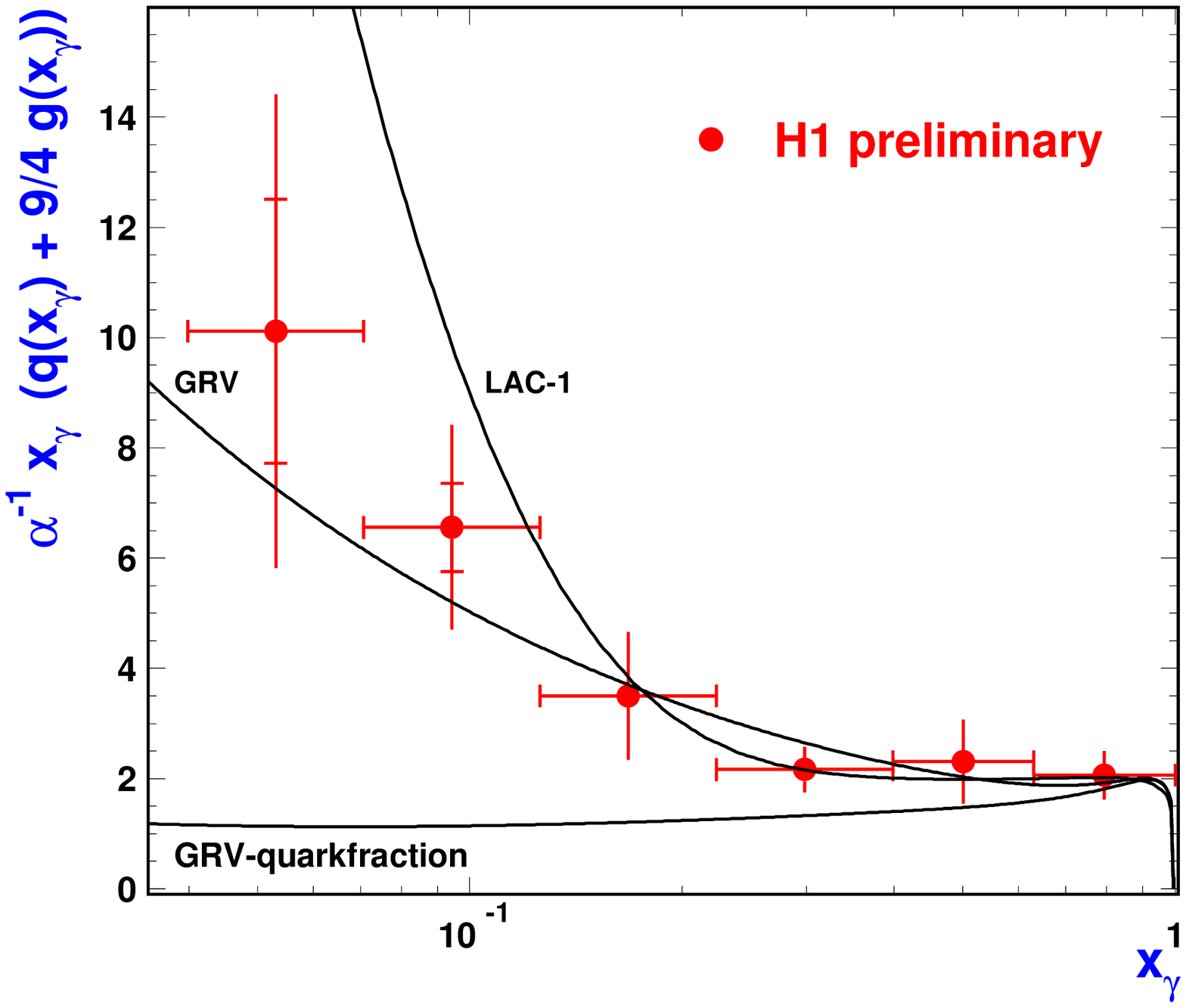}\hspace{.01\hsize}%
\includegraphics[width=.49\hsize,clip]{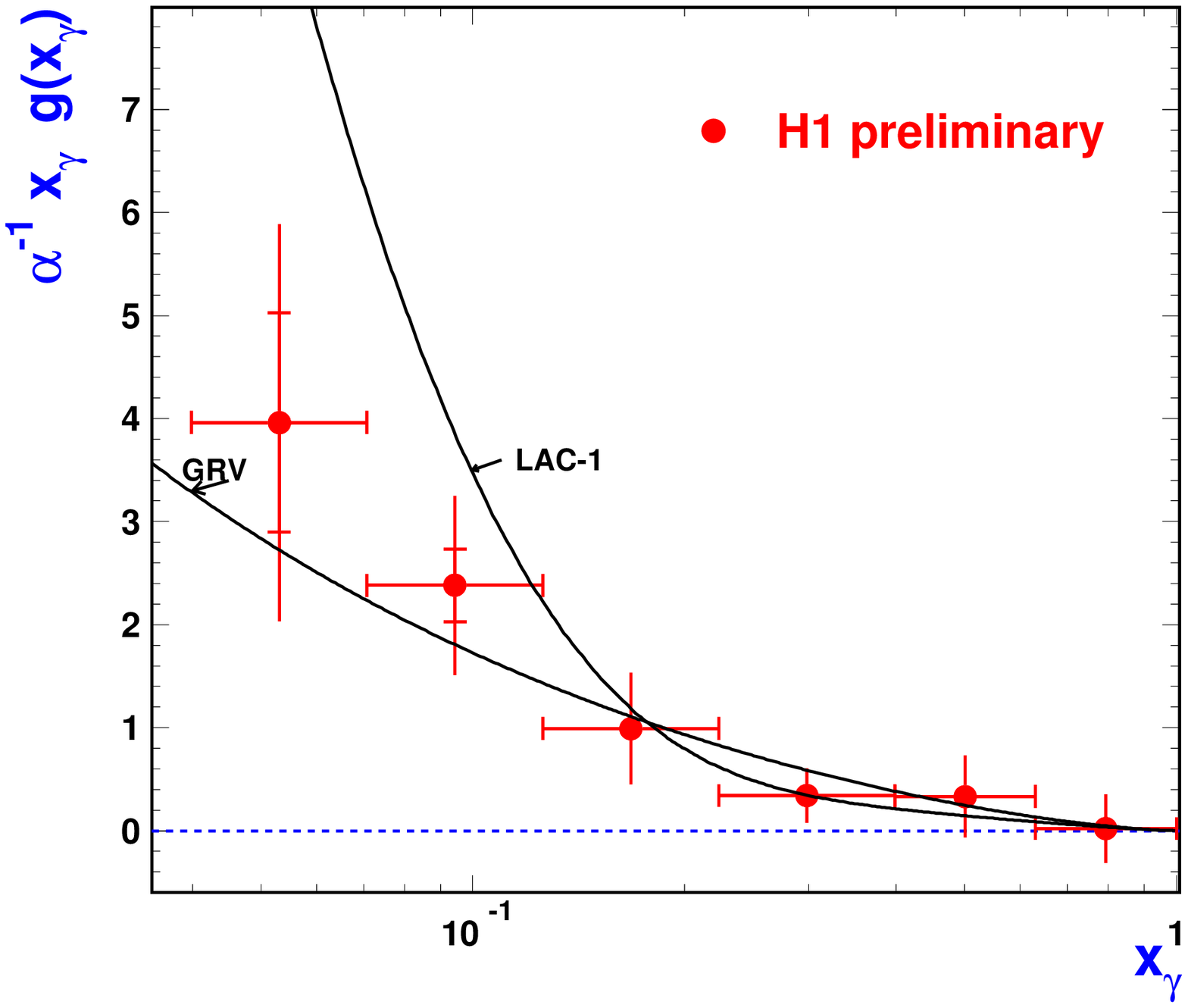}
\caption{Effective photon structure function (left) and gluon density (right)
in the 
photon as extracted from H1 dijet data for photoproduction. For
comparison the LAC-1 and the GRV parameterizations are shown.}
\label{fig:h1:photonstructure}
\end{figure}

The knowledge of the photon structure functions is limited and the pQCD
predictions for the dijet cross sections using different sets vary by 
more than 10\% as can be seen in
figure~\ref{fig:photo:pdf}. H1 extracted an effective photon structure
function from photoproduction\cite{h1photonstructure}. As can be seen in
figure~\ref{fig:h1:photonstructure}, the existence of a gluonic
component in the photon structure is evident.

%{{{  figure
%}}}

%{{{  figure
%}}}
%}}}
%{{{  Summary
\section{Summary}

This article provides a short overview on the variety of measurements of
the hadronic final state in jet production for deep-inelastic scattering
and photoproduction at HERA. Perturbative QCD is successful in
describing the data and therefore allows to extract parameters
such as the strong coupling strength $\alpha_s$ and the parton density
functions with a high precision. An additional reduction of the
uncertainties needs higher order or resumed calculations for the
observables. For photoproduction the importance of the resolved
contribution is clearly demonstrated and a significant measurement of an
effective photon structure function becomes possible. Higher luminosity
will allow to make more precise extractions of the photon structure.

%}}}
%{{{  Acknowledgments
\section*{Acknowledgments}

I would like to thank the organizers for the hospitality and the
invitation to this interesting symposium on a broad variety of topics. I
thank S.~Caron, K.~Rabbertz and P.~Schleper for their discussions and
comments on this article.

%}}}
%{{{  References

%}}}

\begin{thebibliography}{99}

\bibitem{h1det}
 H1 Collaboration,
 %``The H1 detector at HERA,''
 Nucl.\ Instrum.\ Meth.\ {\bf A386}, 310 (1997).
 %%CITATION = NUIMA,A386,310;%%

\bibitem{zeusdet}
 ZEUS Collaboration,
 %``The ZEUS detector: Status report 1993,''
 ZEUS-STATUS-REPT-1993.

\bibitem{ktincl}
  S.D.~Ellis and D.E.~Soper,
  %``Successive combination jet algorithm for hadron collisions,''
  Phys.\ Rev.\ {\bf D48}, 3160 (1993)
  hep-ph/9305266;
  %%CITATION = PHRVA,D48,3160;%%
  \\
  S.~Catani, Y.L.~Dokshitzer, M.H.~Seymour and B.R.~Webber,
  %``Longitudinally invariant K(t) clustering algorithms for hadron-hadron collisions,''
  Nucl.\ Phys.\ {\bf B406}, 187 (1993).
  %%CITATION = NUPHA,B406,187;%%

\bibitem{h1incl}
  H1 Collaboration, 
  ``Determination of the Strong Coupling Constant from Inclusive Jet Cross 
  Sections,'' 
  paper submitted to EPS-HEP99, Tampere, Finland, July 1999.
    
\bibitem{h1disdijet}
  H1 Collaboration,
  ``Measurement of Dijet Cross Sections in Deep Inelastic Scattering at HERA
  and a Direct Determination of the Gluon Density in the Proton,''
  paper submitted to ICHEP98, Vancouver, Canada, July 1998;

\bibitem{zeusdisdijet}
  ZEUS Collaboration,
  ``Measurement of Differential Cross Sections for Dijet Production in Neutral 
  Current DIS at High $Q^2$ and Determination of $\alpha_s$,'' 
  paper submitted to EPS-HEP99, Tampere, Finland, July 1999.

\bibitem{disthree}
  H1 Collaboration,
  ``First Measurement of Three Jet Cross Sections in DIS at HERA,''
  paper submitted to EPS-HEP99, Tampere, Finland, July 1999.

\bibitem{nloprog}
  E.~Mirkes and D.~Zeppenfeld,
  %``Dijet Production at HERA in Next-to-Leading Order,''
  Phys.\ Lett.\ {\bf B380}, 205 (1996)
  hep-ph/9511448;
\\
  S.~Catani and M.H.~Seymour,
  %``A general algorithm for calculating jet cross sections in NLO QCD,''
  Nucl.\ Phys.\ {\bf B485}, 291 (1997)
  hep-ph/9605323;
  %%CITATION = NUPHA,B485,291;%%
\\
  D.~Graudenz,
  %``Deeply inelastic hadronic final states: QCD corrections,''
  hep-ph/9708362;
  %%CITATION = HEP-PH 9708362;%%
\\
  B.~P{\"o}tter,
  %``JetViP 1.1: Calculating one- and two-jet cross sections with virtual  photons in NLO QCD,''
  Comput.\ Phys.\ Commun.\ {\bf 119}, 45 (1999)
  hep-ph/9806437.
  %%CITATION = CPHCB,119,45;%%

\bibitem{photodijet:eps}
  ZEUS Collaboration,
  ''Measurement of Dijet Photoproduction at High Transverse Energies at HERA,''
  paper submitted to EPS-HEP99, Tampere, Finland, July 1999;

\bibitem{photodijet:paper}
  ZEUS Collaboration,
  %``Measurement of dijet photoproduction at high transverse energies at  HERA,''
  hep-ex/9905046.
  %%CITATION = HEP-EX 9905046;%%
  
\bibitem{photothree}
  ZEUS Collaboration,
  ``Three-jet Distributions in Photoproduction at HERA,''
  paper submitted to EPS-HEP99, Tampere, Finland, July 1999.
  
\bibitem{h1photonstructure}
  H1 Collaboration,
  ``Dijet Cross Sections in Photoproduction and Determination of the Gluon
  Density in the Photon,'' 
  figures submitted to EPS-HEP99, Tampere, Finland, July 1999.

\end{thebibliography}
\end{document}